\documentclass[pra,floatfix,amsmath,twocolumn,showpacs,superscriptaddress]{revtex4}
\usepackage[ansinew]{inputenc}

\usepackage{verbatim}
\usepackage{epsfig}
\usepackage{pst-all}
\usepackage{color}
\usepackage{dcolumn}
\usepackage{amsmath}
\usepackage{amsthm}
\usepackage{bm}
\usepackage{layout}
\usepackage{float}
\usepackage{txfonts}
\usepackage{amsfonts}
\usepackage{amssymb}%
\usepackage[ansinew]{inputenc}
\usepackage{amsmath}
\usepackage{amsthm}
\usepackage{float}
\usepackage{amsfonts}
\usepackage{amssymb}

\begin{document}

\title{Classification of no-signaling correlations
and the guess your neighbor's input game}

\author{He-Ming Wang}
\affiliation{School of Physics, Peking University, Beijing 100871, China}
\author{Heng-Yun Zhou}
\affiliation{School of Physics, Peking University, Beijing 100871, China}
\affiliation{Department of Physics, MIT, Cambridge, MA 02139, USA}

\author{Liang-Zhu Mu}
\email{muliangzhu@pku.edu.cn}
\affiliation{School of Physics, Peking University, Beijing 100871, China}


\author{Heng Fan}
\email{hfan@iphy.ac.cn}
\affiliation{Institute of Physics, Chinese Academy of Science, Beijing 100190, China}
\affiliation{Collaborative Innovation Center of Quantum Matter, Beijing 19910, China}


\date{\today}

\begin{abstract}
We formulate a series of non-trivial equalities which are satisfied by
all no-signaling correlations, meaning that no faster-than-light
communication is allowed with the resource of these correlations.
All quantum and classical correlations satisfy
these equalities since they are no-signaling. By applying these equalities, we provide a general
framework for solving the multipartite ``guess your neighbor's input'' (GYNI) game,
which is naturally no-signaling but shows conversely that general no-signaling correlations are actually more non-local than those
allowed by quantum mechanics. We
confirm the validity of our method for number of players from 3 up to 19,
thus providing convincing evidence that it works for the general case.
In addition, we solve analytically the tripartite GYNI
and obtain a computable measure of supra-quantum correlations.
This result simplifies the defined optimization procedure
to an analytic formula, thus
characterizing explicitly the boundary between quantum and supra-quantum correlations.
In addition, we show that the gap between quantum and no-signaling boundaries
containing supra-quantum correlations can be closed
by local orthogonality conditions in the tripartite case.
Our results provide a computable classification of no-signaling correlations.
\end{abstract}


\maketitle

\section{Introduction}

Quantum mechanics allows non-local correlations such
as the Einstein, Podolsky, and Rosen (EPR) pairs \cite{EPR}.
Entanglement like EPR pairs can be used
as a valuable resource for quantum information processing \cite{Nielsen&Chuang,ekert} such as
the well-known quantum teleportation \cite{teleportation,teleportation-experiment,teleportation-experiment1}.
However, quantum teleportation relies on classical communication
for state transmission and thus will not violate
the no-signaling condition, meaning that signals cannot be sent faster-than-light.
In fact, no-signaling is a general principle of quantum mechanics and
it is thus satisfied by all non-local quantum correlations.
It is also closely related but different from quantum causality \cite{BancalNP,Causality,brukner}.
A broad class of theories exist which can characterize the nonlocality of
quantum physics, such as the Bell inequalities \cite{Bell,Bellnonlocality} and
the temporal analogue Leggett-Garg inequality \cite{leggett-garg}, see also
results in Ref.\cite{tsirelson}.
In particular, due to the recent advent of quantum information,
the extremely intense study of quantum correlations such as
entanglement and discord has made nonlocality widely appreciated as a
fundamental property of various quantum systems,
see \cite{eof,hv01,oz02,Modirmp,horodeckirmp,Amico-RMP,Cui-Gu,LPSW,NavascuesPRL,BarnumPRL10,Unified10}
and the references therein for related topics.

On the other hand, it is known that conversely there exist no-signaling correlations more
nonlocal than those allowed in quantum mechanics \cite{PRbox}, see a recent review
paper by Popescu and the references therein \cite{popescu}.
Recently, a nonlocal multipartite scheme GYNI, ``guess your neighbor's input'', has been presented and investigated in
Refs. \cite{GYNI,winternature,GYNIbook,Fritz,GallegoPRL,YangTHNJP,PironioJPA,AugusiakPRA,UPB-PRL11}. It demonstrates that
the no-signaling correlations provide a clear advantage over both classical and
quantum correlations, while these two correlations have a common ground in this scheme.
This scheme leads to a facet Bell inequality which is true for quantum correlations
and is not implied by any other Bell inequalities.
Yet its violation is consistent with no-signalling,
see a views paper \cite{winternature} for
the implications and importance of the GYNI scheme.
Despite the significant role of GYNI in clarifying the concepts of quantum correlations and fundamentals of quantum mechanics, the
scheme itself is largely unsolved, even for the simplest tripartite scenario.
The optimal advantage of no-signaling in GYNI has been demonstrated analytically for $N=3$, numerically for $N=5,7$ cases,
under the assumption of a given probability distribution.
For years, with much progress and understandings related to this
game, the solution of the GYNI game still seems challenging.
By studying the GYNI scheme, we can distinguish quantum correlations from
other supra-quantum no-signaling correlations and find the borderline between them.
We can also explore the upper boundary of all no-signaling correlations.
The parallel situation is
the Bell inequality which can distinguish quantum correlations from classical correlations
and can also be used to explore the upper bound of quantum correlations.
Besides, the GYNI scheme may have important implications in understanding
quantum physics and information theory.

In this paper, we propose and formulate a series of
non-trivial equalities.
These equalities capture the common properties of
no-signaling correlations in a precise way.
Based on these equalities, a general framework to solve the GYNI
problem is provided.
We confirm the validity of our solution for a number of players from $N=3$ up
to $N=19$. This provides convincing evidence that the framework works.
We show that the advantage of no-signaling correlations over quantum or classical
cases scales to the proven bound 2 \cite{GYNI,GYNIbook} and the correlations
achieving the optimal bound are given.
Additionally, we solve analytically the tripartite case completely.
A concise form of the winning probability ratio
between no-signaling and classical or quantum correlations is obtained,
which is computable analytically and thus avoids the optimization procedure.
This identifies clearly the boundary between quantum
correlations and no-signaling supra-quantum correlations.
We also notice that with the local orthogonality condition \cite{Fritz},
the gap harboring the existence of supra-quantum correlations can be closed
in the case of the tripartite system. This fact confirms the
necessity of local orthogonality in the GYNI game for quantum mechanics.

\section{GYNI and the no-signaling equalities}

Let us begin with the game of GYNI \cite{GYNI} shown in FIG.1.
A number $N$ of players are in a round-table meeting and each receives
a poker of `heart' or `spade' representing input bit $x_i\in \{0,1\}$.
The aim is that each player provides an output bit
$a_i\in \{0,1\}$ representing the guess about his/her
right-hand neighbor's input.
No communication is allowed after the inputs are distributed and thus
no-signaling is ensured.
The input strings ${\bf x}=x_1,...,x_N$
are chosen according to some prior fixed probability distributions
$q({\bf x})$ known to all players,
where $P(a_1a_2...a_N|x_1x_2...x_N)$ is the probability of obtaining the output
$a_1a_2...a_N$ when the input $x_1x_2...x_N$ is given.
The probabilities satisfy the identity $P(x,...,x|0,...,0)=1$,
meaning the probability summation over
all possible outputs for a given input is 1, where `x' is assumed to
be a summation of all possible outputs at each position.
The correct output probability is denoted as $P({\bf a_i}={\bf x}_{i+1}|{\bf x})$.
The average winning probability is thus quantified as
$\omega =\sum _{\bf x}q({\bf x})P({\bf a_i}={\bf x}_{i+1}|{\bf x})$.
The winning probabilities by classical strategies and general no-signaling ones
are denoted as $\omega _c,\omega _{ns}$, respectively.
No quantum advantage over the classical case is available in this game \cite{GYNI},
meaning that $\omega _c$ is applicable for the quantum case. This is due to the
condition that no communication is allowed in this scheme.
We remark that the $N=2$ case is trivial.

\begin{figure}
\includegraphics[width=8cm]{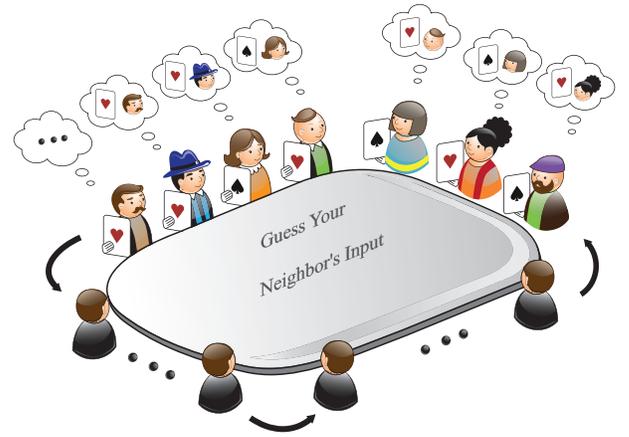}
\caption{(color online) The game of ``guess your neighbor's input'' \cite{GYNI}.
The aim is that each player provides an output bit
$a_i\in \{0,1\}$ representing the guess about his/her
right-hand neighbor's input. Here $\{0,1\}$ are represented by
`spade' and `heart' of the pokers.
No communication is allowed after the inputs are distributed in this game.}
\label{roundtable}
\end{figure}

We study the GYNI game first by considering the odd $N$ case and the specified input distribution,
$q({\bf x})=1/2^{N-1}$ when $x_1\oplus x_2\oplus ...\oplus x_N=0$ and $q({\bf x})=0$ otherwise,
which we stick to in this work unless stressed explicitly.
We know that $\omega_c=1/2^{N-1}$ for both classical and quantum resources \cite{GYNI}.
Now we show that for no-signaling resources,
\begin{equation}
\label{oddspecial}
\max \omega_{ns}=\frac{2}{1+C_{N-1}^{\frac{N-1}{2}}/2^{N-1}}\omega_c.
\end{equation}
We can directly verify that $\max \omega _{ns}/\omega _c$ is larger than 1 and
scales to 2 for large $N$, thus saturating the upper bound \cite{GYNI},
see FIG.2. Hereafter, we generally explore the upper bound of $\omega _{ns}$
and the notation `$\max $' will be dropped with no confusion.

Our proof of Eq.(\ref{oddspecial}) is based on a series of no-signaling equalities presented below.
These equalities belong to facet Bell inequalities, meaning that they are not
violable by quantum mechanics and are not implied by other Bell inequalities.
On the other hand, they are equalities instead of inequalities and we may name them
as {\it Bell equalities}, see FIG.3 for explanations about their role in classifying no-signaling correlation.
Our proposed no-signaling Bell equalities for the concerned
probabilities in the GYNI game take the form:
\begin{equation}
\label{equality}
\sum_{\sum x_i<\frac{N}{2}}(P(x_2...x_Nx_1|x_1x_2...x_N)+P(x_2'...x_N'x_1'|x_1x_2...x_N))=1,
\end{equation}
where the summation is under the condition
that the sum of $x_i$ is less than $N/2$. The first half terms $P(x_2...x_Nx_1|x_1x_2...x_N)$ are of GYNI interest, and the second half of the terms $P(x_2'...x_N'x_1'|x_1x_2...x_N)$ are the
pairing terms corresponding to the first half. The one-to-one correspondence of $x_1'x_2'...x_N'$ and $x_1x_2...x_N$ is given by
\begin{equation}
x_i'=\left\{
\begin{array}{ll}
0&\mathrm{if\ }x_i=0\mathrm{\ and\ }\exists j,\ \sum_{k=1}^j(2x_{i+k}-1)>0\\
1&\mathrm{otherwise.}
\end{array}
\right.
\end{equation}
Intuitively, this construction starts from the terms with the most 1's and changes one of the $x_i$'s from 1 to 0; it then cascades down until all probability terms have $x_1x_2...x_N=\mathbf{0}$ so that normalization conditions can be used. The proof of the Bell equality is due to the no-signaling principle
$P(0x|01)=P(0x|00)$ and the identity $P(x,...,x|0,...,0)=1$.
Detailed discussion of the proof and the correspondence can be found in Appendix A.

Now we turn to the upper bound for no-signaling GYNI winning probabilities
which are in the first part of Eq.(\ref{equality}).

With $N=5$ as a simple example, the 11 terms with zero or two 1's appear in the equality (\ref{equality}) just proven, so that the sum of these terms is less than or equal to 1. By relabeling inputs and outputs using the 0,1 symmetry and maintaining an even number of 1's to match the terms of GYNI interest, we can find a total of 16 similar inequalities, with each term in the expression of $\omega_{ns}$ appearing 11 times due to relabeling symmetry. This then gives the upper bound $\omega_{ns}|_{N=5}\leq \frac{1}{16}\times 16/11=\frac{1}{11}$.

Similar to above, combinatorial considerations let us collect the terms containing $0,2,4...2m$ 1s in the case of $N=4m+1$, or $4m+2,4m,4m-2...2m+2$ 1s in the case of $N=4m+3$, where $m$ is a positive integer.
The number of these terms, as a function of odd $N$, can be expressed as $\sum_{i=0}^m C_{4m+1}^{2i}$ or $\sum_{i=0}^m C_{4m+3}^{2m+2+2i}$, which can both be reduced to $2^{N-2}+C_{N-1}^{(N-1)/2}/2$ using the combinatorial relation $C_{m+1}^{n+1}=C_m^n+C_m^{n+1}$.
This shows the upper bound of the no-signaling winning probability,
\begin{equation}
\label{upperbound}
\omega_{ns}\leq\frac{2}{2^{N-1}+C_{N-1}^{\frac{N-1}{2}}}.
\end{equation}

\section{The tight upper bound and the constituents of no-signaling correlations}

Now we demonstrate that no-signaling correlations saturating the inequality (\ref{upperbound}) can be found.
Thus the inequality turns out to be an equality in the optimal case.
Generally the number of inequalities is less than the degrees of freedom of the correlations, and such correlations are not unique. However, under the restrictions of basic symmetry (invariance under relabeling inputs, outputs and states), only one solution can be found for $N=3$, 5 and 7. For $N=9$, the solutions have two degrees of freedom, and when $N$ becomes larger the degrees of freedom increase exponentially.

The only no-signaling symmetric correlation for $N=3$ can be expressed as:
\begin{eqnarray}
\label{extremeNS3}
P(abc|xyz)&=&\frac{1}{6}(x\oplus y\oplus z\oplus xy\oplus yz\oplus zx\oplus ab\oplus bc\oplus ca
\nonumber \\
&&\oplus zb\oplus ya\oplus xc)+\frac{1}{3}\bar{a}\bar{b}\bar{c}\bar{x}\bar{y}\bar{z}+\frac{1}{3}a\bar{b}cxy\bar{z}
\nonumber \\
&&+\frac{1}{3}\bar{a}bcx\bar{y}z+\frac{1}{3}ab\bar{c}\bar{x}yz.
\end{eqnarray}

In these correlations the GYNI probability terms, $P(x_2...x_Nx_1|x_1x_2...x_N)$ with $x_1\oplus x_2\oplus ...\oplus x_N=0$,
are all equal to $2/(2^{N-1}+C_{N-1}^{(N-1)/2})$,
while the Bell equality (\ref{equality}) can be satisfied.
This means that the winning probabilities achieve the upper bound.
The advantage of no-signaling over quantum and classical
correlations takes the form (\ref{oddspecial}). We remark that this
result is confirmed for a number of players up to 19 by
a computer workstation (16-core, 384G-memory).
The calculations involve the proof of the Bell
equality (\ref{equality}) and the saturating of the
bound (\ref{upperbound}) both for $N$ up to 19.

FIG. 2 shows the asymptotical behavior of the winning ratio of no-signaling
correlations over quantum or classical correlations (Eq. (\ref{oddspecial})).

\begin{figure}
\includegraphics[width=5cm]{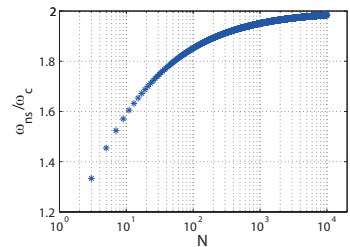}
\caption{(color online) Winning probability ratio
for no-signaling correlations over classical or quantum correlations.
We assume $N$ is odd.
The star symbols represent
our result of Eq.(\ref{oddspecial}),
which will approach 2 asymptotically when $N$ is large.
 }
\label{compare}
\end{figure}

\begin{figure}
\includegraphics[width=8cm]{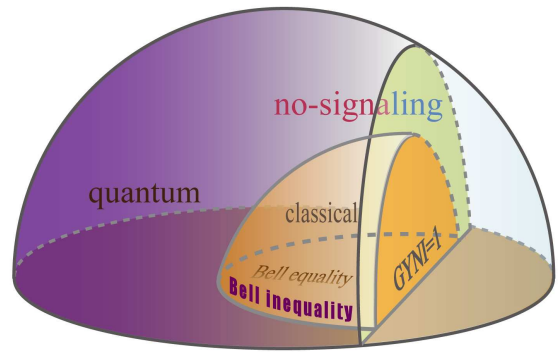}
\caption{(color online) Schematic representation of various
correlations. Those correlations include classical, quantum and
no-signaling correlations. The Bell equality and the GYNI game can identify
different boundaries between those correlations. The well-known
Bell inequality is also marked.
}
\label{relation}
\end{figure}

FIG. 3 shows the above proposed Bell equality and
GYNI game in describing various correlations.
The largest volume in the figure represents no-signaling
correlations. The no-signaling correlations contain quantum correlations
as a subset, while the quantum correlations contain classical
correlations as a subset. They all share a partly common boundary
which is the bottom of the classical volume in this figure.
It is described by our proposed Bell equality and thus is marked as `Bell equality'.
The Bell equality can be checked for general tripartite qubit state
which is proven to take a simple form in Ref. \cite{triquantumbit}.
The violation of the Bell equality means that no-signaling correlations cannot
accommodate this phenomenon.
Quantum correlations and classical correlations may share a
common part of boundary distinguishing them from
supra-quantum correlations which satisfy the no-signaling condition but
are beyond quantum mechanics. This boundary can
be identified by $\omega _{ns}/\omega _c=1$ which is marked as `GYNI=1' in this figure.
For the tripartite case,
we can identify this boundary analytically by using Eq.(\ref{triboundary}) presented later
when it equals to 1. This is the first computable measure of supra-quantum correlations.
The $\omega _{ns}/\omega _c>1$ part belongs solely to no-signaling supra-quantum correlation with the
boundary identified by Eq.(\ref{oddspecial}).
The well-known Bell inequality distinguishes
quantum correlations and classical correlations which are also marked in this figure.

\section{Analytic formula, no-signaling inequalities and local orthogonality}

We next consider arbitrary given inputs and assume
$N\geq3$ being both odd and even numbers.
The necessary probability inequalities in the tripartite case come from various Bell equalities
and have a clear geometric representation.

FIG. 4 shows those inequalities for the tripartite case.
A total of 14 restrictions on the probability terms can be found using the hypercube-hyperplane representation of inequalities, which can be divided into two classes. Left panel (a) represents
the first class of 6 inequalities that can be represented by $P(000|000)+P(010|001)+P(001|100)+P(011|101)\leq 1$ which corresponds to the normal vector $(0,1,0)$.
Right panel (b) represents the second class of 8 inequalities that can be represented by $P(000|000)+P(010|001)+P(100|010)+P(001|100)\leq 1$, which corresponds to the normal vector $(1,1,1)$.
In general, we can construct a hypercube of dimension $N$ in a Cartesian space
and all the vertices of the hypercube have a coordinate $(x_1,x_2,...x_N)$ corresponding
to GYNI interest probability $P(x_2...x_Nx_1|x_1x_2...x_N)$.
We propose that, for every hyperplane passing through the center of the $N$-dimensional hypercube while not passing through any vertex, there exists a pair of corresponding inequalities that limit the sums of the probability terms on the two sides of the hyperplane.
Explicitly, given a real vector ${s_1,s_2,...s_N}$, if the equations $\sum_ix_is_i-N/2=0$ and $x_i\in\{0,1\}$ have no solutions then we should have
$\sum_{\sum_i x_ia_i<\frac{N}{2}}P(x_2...x_Nx_1|x_1x_2...x_N)\leq 1.$
This has been proven to be true for $N=3,4$ and $5$ by direct verification.
Instead of the original no-signaling principles,
these inequalities can be used to estimate the upper bound of $\omega_{ns}$.
However, we remark that they are not complete for general $N$.
Using these inequalities, we can find that, for a given input distribution $q(\mathrm{\textbf{x}})$, the maximum ratio of winning probability is
\begin{eqnarray}
&&\max \frac{\omega_{ns}}{\omega_c}=\max (1,\frac{q(000)+q(110)+q(101)+q(011)}{3\omega_c},
\nonumber \\
&&~~~~~~~~~~~~~~~~\frac{q(100)+q(010)+q(001)+q(111)}{3\omega_c}),
\label{triboundary}
\end{eqnarray}
where $\omega_c=\max (q(\mathrm{\textbf{x}})+q(\mathrm{\bar{\textbf{x}}}))$ is the classical winning probability (see Appendix B for a detailed proof).
This is the first analytic measure of supra-quantum correlations without optimization.
\begin{figure}
\includegraphics[width=8cm]{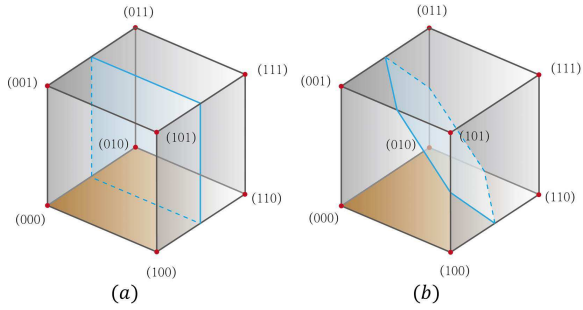}
\caption{(color online) Geometric representation of no-signaling inequalities for the tripartite case.
Cases for larger $N$ can also be represented in a similar way.
Two classes of inequalities corresponding respectively to left and right panels are
presented.
}
\label{cube}
\end{figure}

Furthermore, the ratio derived here for arbitrary input distributions can always be reached. Consider two no-signaling resources: the first one is given by $P(abc|xyz)=(a\oplus y\oplus x\oplus x')(b\oplus z\oplus y\oplus y')(c\oplus x\oplus y\oplus y')$, the classical strategy described in \cite{GYNI}, where $\mathrm{\textbf{x}}=x'y'z'$ maximizes $(q(\mathrm{\textbf{x}})+q(\mathrm{\bar{\textbf{x}}}))$,
and the second one is the extreme no-signaling correlation (\ref{extremeNS3}). It can be easily checked that after suitable relabeling, the first one gives $\omega_{ns}=\omega_c$, and the second one gives $3\omega_{ns}=\max(q(000)+q(110)+q(101)+q(011),q(100)+q(010)+q(001)+q(111))$.

We can also look at the no-signaling GYNI game with four parties for arbitrary input distributions.
We first show that no input distributions can achieve a higher probability ratio than the distributions satisfying $q(\textbf{x})\in\{0,1/2^{N-1}\}$ and $q(\textbf{x})+q(\bar{\textbf{x}})=1/2^{N-1}$. From \cite{GYNI}, we know that $\omega_c=\max_{\textbf{x}} q(\textbf{x})
+q(\bar{\textbf{x}})$.
If for some ${\bf x}$, $q(\textbf{x})+q(\bar{\textbf{x}})<\omega_c$, then we can increase $q(\textbf{x})$
so that $q(\textbf{x})+q(\bar{\textbf{x}})=\omega_c$ and renormalize $q(\textbf{x})$.
In this process $\omega_{ns}$ does not decrease and $\omega_c$ does not change, so the ratio does not decrease.
When $q(\textbf{x})+q(\bar{\textbf{x}})=\omega_c$ for all $\textbf{x}$, we must have $\omega_c=1/2^{N-1}$.
Since $\omega_{ns}$ is linear in $q(\textbf{x})$ for a fixed no-signaling correlation,
the maximum can only be achieved when either $q(\textbf{x})=0,\ q(\bar{\textbf{x}})=1/2^{N-1}$
or $q(\bar{\textbf{x}})=0,\ q(\textbf{x})=1/2^{N-1}$.

By the above reasoning, we only need to check a small number of input distributions.
This then becomes some linear programming problems with several target functions.
There are three inequivalent normal vectors, $(1,0,0,0)$, $(5,2,2,2)$, and $(1,1,1,0)$ in the
hypercube-hyperplane (geometric) representation, corresponding to a total of 104 inequalities.
Under these restrictions, we tested all 256 input distributions in the form $q(\textbf{x})\in\{0,1/8\}$ and $q(\textbf{x})+q(\bar{\textbf{x}})=1/8$.
Calculations show that
$\max \omega_{ns}/\omega_c|_{N=4}=4/3$. The result shows that no matter how the input distribution is given, adding a party into the game cannot help no-signaling resources doing better.

Remarkably, it has been recently proposed that local orthogonality may play a critical role for
the characterization of quantum mechanics, in particular for distinguishing supra-quantum correlations \cite{Fritz}.
Explicitly, local orthogonality offers new inequalities satisfied by the correlations. By some
calculations (presented in Appendix B), we find that the gap between boundaries of no-signaling and quantum in GYNI which
harbors supra-quantum correlation will be completely closed by local orthogonality conditions. This fact confirms that
local orthogonality is necessary in distinguishing supra-quantum correlation.

\section{Conclusions}

In summary, we provide a valid framework for solving
the general GYNI game by introducing a series of non-trivial equalities which
might be named as Bell equalities.
All no-signaling correlations satisfy these equalities and
the violation of them means the violation of no-signaling.
We also obtain a concise form for measure of
supra-quantum correlations without relying on an optimization procedure which is generally
a hard task. We remark that the supra-quantum correlations will be removed by local orthogonality conditions.
Our results offer a classification of no-signaling correlations.
A lot of new questions arise related to the results in this article. For example,
the proof of general Bell equalities and their applications, measures of supra-quantum
correlation for more general cases, the explicit relationship between Bell equality
and Bell inequality. These are still open problems which are worth studying further.

\begin{acknowledgments}
This work was supported by the `973' Program
(2010CB922904), NSFC (11175248), NFFTBS (J1030310, J1103205) and
grants from the Chinese Academy of Sciences.
\end{acknowledgments}

\appendix

\section{Proof and the correspondence of the no-signaling equality}

We first present the proof for $N=5$. We write the equality explicitly and get:
\begin{eqnarray}
&&P(00000|00000)+\sum_c( P(10000|01000)
\nonumber \\
&&+P(11000|01100)+P(10100|01010)
\nonumber \\
&&+P(10110|01010)+P(11100|01100)
\nonumber \\
&&+P(11110|01000))+P(11111|00000)=1,
\end{eqnarray}
where the subscript `c' means cyclic summation of the input and output.

To prove this equality, note that
\begin{eqnarray}
&&P(10000|01000)+P(11000|01100)\nonumber \\
&&+P(10100|01010)+P(10110|01010)\nonumber \\
&&+P(11100|01100)+P(01111|01000)\nonumber \\
&=&P(10000|01000)+P(11x00|01000)\nonumber \\
&&+P(101x0|01000)+P(11110|01000)\nonumber \\
&=&P(1xx00|00000)+P(1x110|00000),
\end{eqnarray}
where an `x' in the output stands for summation of all possible states at that position, and the no-signalling principle has been used several times. Substitute this into the left hand side and the equation can be transformed into:
\begin{eqnarray}
&&P(00000|00000)+\sum_c(P(1xx00|00000)\nonumber \\
&&+P(1x110|00000))+P(11111|00000)\nonumber \\
&=&P(xxxxx|00000)=1,
\end{eqnarray}
which completes the proof of this equality for $N=5$. Proofs for larger $N$s can be constructed similarly with the aid of a computer.

The correspondence between the pair $x_1'x_2'...x_N'$ and $x_1x_2...x_N$ was described as an existence criterion. Now we give a different way of constructing the correspondence. Consider the following procedures:

\begin{enumerate}
\item Copy the input string onto a piece of cyclic paper (which ensures that $x_{N+1}$ is equivalent to $x_1$).
\item For each `1' on the paper, cross it out, then find the nearest `0' not crossed out on the left of this `1' and cross this `0' out.
\item For the numbers not crossed-out (which must be `0'), change them to `1'.
\item Now the paper contains a new string, which is the desired $x_1'x_2'...x_N'$.
\end{enumerate}

We have to clarify some points of this procedure. First, it should be obvious that step 2 of this procedure is well-defined; that is, the result is the same regardless of the sequence we cross out 1s. Second, it is always possible to finish step 2, since for $x_i$ we have the restriction $\sum x_i<\frac{N}{2}$, and there are more zeros than ones in the input string. Each one `cancels' a zero, and there should be an equal number of crossed-out zeros and ones.

We now show that the procedure is equivalent to the existence criteria we have given. The statement `for some j $\sum_{k=1}^j(2x_{i+k}-1)>0$' is equivalent to saying that `for some j there are more ones than zeros in $x_{i+1}..x_{i+j}$'. By step 2 of the procedure we described, the zero at this position will be `canceled' by a one from its right and remain as a zero in its pair string. If this is not the case, then either $x_i=1$, or $x_i=0$ which is not crossed out and $x_i'=1$.

Using this procedural description, we can also show that this correspondence is one-to-one. Given $x_1'x_2'...x_N'$, we can go through a similar procedure in the reverse way to recover the original string $x_1x_2...x_N$. For all `0's in the string, cross it out, and find the nearest `1's not crossed out on the right of this `0' and cross them out. Then the remaining `1's should be changed to `0's.

We notice that the pairing process always introduce new terms satisfying $\sum_{i=1}^N(x_i+x_i')=N$. We believe that this pairing process should have physical implications, but by now we have not found a clear explanation of this.

\section{Derivation of the maximum ratio}

Here we prove the maximum ratio of winning probability for $N=3$, expressed as a maximizing function of the input distribution.

For simplicity, we introduce some short-form notations: $P(000|000)$ will be shortened to $P_0$, $P(010|001)$ to $P_1$, $q(010)$ to $q_2$, and so on. Then the maximum ratio becomes:

\begin{equation}
\max \frac{\omega_{ns}}{\omega_c}=\max (1,\frac{q_0+q_3+q_5+q_6}{3\omega_c},\frac{q_1+q_2+q_4+q_7}{3\omega_c}),
\end{equation}
where $\omega_c=\max (q_0+q_7,q_1+q_6,q_2+q_5,q_3+q_4)$ is the classical winning probability.

The proof is divided into two parts. The first part assumes $q_0+q_3+q_5+q_6\geq3\omega_c$. We start by choosing 4 inequalities out of 14, namely
\begin{eqnarray}
P_0+P_1+P_3+P_5\leq1 \nonumber \\
P_0+P_2+P_3+P_6\leq1 \nonumber \\
P_0+P_4+P_5+P_6\leq1 \nonumber \\
P_3+P_5+P_6+P_7\leq1
\end{eqnarray}

We multiply these 4 inequalities by $q_0+q_3+q_5-2q_6$, $q_0+q_3+q_6-2q_5$, $q_0+q_5+q_6-2q_3$ and $q_3+q_5+q_6-2q_0$ respectively. Since $q_0+q_3+q_5+q_6\geq3\omega_c\geq3q_1+3q_6$, $q_0+q_3+q_5-2q_6\geq3q_1\geq0$ and by symmetry all the coefficients here are nonnegative. Adding these together and we get

\begin{eqnarray}
3(q_0P_0+q_3P_3+q_5P_5+q_6P_6) \nonumber \\
+(q_0+q_3+q_5-2q_6)P_1+(q_0+q_3+q_6-2q_5)P_2 \nonumber \\
+(q_0+q_5+q_6-2q_3)P_4+(q_3+q_5+q_6-2q_0)P_7 \nonumber \\
\leq q_0+q_3+q_5+q_6.
\end{eqnarray}

Since $q_0+q_3+q_5-2q_6\geq3q_1$, we have $(q_0+q_3+q_5-2q_6)P_1\geq3q_1P_1$, thus changing the left hand side gives
\begin{eqnarray}
3(q_0P_0+q_3P_3+q_5P_5+q_6P_6+q_1P_1+q_2P_2 \nonumber \\
+q_4P_4+q_7P_7)\leq q_0+q_3+q_5+q_6.
\end{eqnarray}

Noticing that $\omega_{ns}=\sum_{i=0}^{7}q_iP_i$, we have
\begin{equation}
\omega_{ns}\leq \frac{q_0+q_3+q_5+q_6}{3}
\end{equation}

The proof for $q_1+q_2+q_4+q_7\geq3\omega_c$ is similar and we have $\omega_{ns}\leq (q_1+q_2+q_4+q_7)/3$.

The second part assumes $q_0+q_3+q_5+q_6<3\omega_c$ and $q_1+q_2+q_4+q_7<3\omega_c$. We prove that $\omega_{ns}/\omega_c\leq1$ by constructing a new input distributions so that $\omega'_{ns}/\omega'_c\geq\omega_{ns}/\omega_c$ and then show that $\omega'_{ns}\leq\omega'_c$.

$q'$ is constructed from $q$ as follows:
\begin{eqnarray}
Q'_i=q_i+(\omega_c-q_i-q_{7-i})s_i/(s_i+s_{7-i}) \nonumber \\
q'_i=Q'_i/\sum_{j=0}^7Q'_j,
\end{eqnarray}
where

\begin{eqnarray}
s_0=s_3=s_5=s_6=3\omega_c-q_0-q_3-q_5-q_6 \nonumber \\
s_1=s_2=s_4=s_7=3\omega_c-q_1-q_2-q_4-q_7
\end{eqnarray}

The construction increases each $q_i$ to $Q'_i$ so that $Q'_i+Q'_{7-i}$ are equal for all $i$, while keeping $Q'_0+Q'_3+Q'_5+Q'_6\leq3\omega_c$ and $Q'_1+Q'_2+Q'_4+Q'_7\leq3\omega_c$. After this adjusting $Q'_i$ is normalized to give $q'_i$. It's easy to see that $q'_i+q'_{7-i}=\omega'_c$ for all $i$ and

\begin{equation}
\omega'_{ns}/\omega'_c=\sum_{i=0}^{7}Q'_iP_i/\omega_c\geq\omega_{ns}/\omega_c.
\end{equation}

Now without loss of generality, we assume that $\min q'_i=q'_7$. Because $Q'_0+Q'_3+Q'_5+Q'_6\leq3\omega_c$, $q'_0+q'_3+q'_5+q'_6\leq3\omega'_c$ and $q'_0\leq q'_1+q'_2+q'_4$.

If $(q'_1+q'_2+q'_4-q'_0)/2\leq\min (q'_1,q'_2,q'_4)$, we use the following inequalities:

\begin{eqnarray}
(q'_1+q'_2+q'_4-q'_0)/2 \times (P_0+P_1+P_2+P_4\leq1) \nonumber \\
(q'_1+q'_0-q'_2-q'_4)/2 \times (P_0+P_1+P_3+P_5\leq1) \nonumber \\
(q'_2+q'_0-q'_1-q'_4)/2 \times (P_0+P_2+P_3+P_6\leq1) \nonumber \\
(q'_4+q'_0-q'_1-q'_2)/2 \times (P_0+P_4+P_5+P_6\leq1) \nonumber \\
(q'_7) \times (P_3+P_5+P_6+P_7\leq1) \nonumber \\
\end{eqnarray}

It can be easily verified that all coefficients are non-negative. Adding all of these together and we get
\begin{eqnarray}
q'_0P_0+q'_1P_1+q'_2P_2+q'_4P_4+(q'_0+q'_7-q'_4)P_3 \nonumber \\
+(q'_0+q'_7-q'_2)P_5+(q'_0+q'_7-q'_1)P_6+q'_7P_7 \nonumber \\
\leq q'_0+q'_7.
\end{eqnarray}

Similarly, if $(q'_1+q'_2+q'_4-q'_0)/2\geq\min (q'_1,q'_2,q'_4)$, without the loss of generality we set $\min (q'_1,q'_2,q'_4)=q'_1$, then we use the following inequalities:

\begin{eqnarray}
q'_1                  \times (P_0+P_1+P_2+P_4\leq1) \nonumber \\
(q'_2+q'_4-q'_1-q'_0) \times (P_0+P_2+P_4+P_6\leq1) \nonumber \\
(q'_0-q'_4) \times (P_0+P_2+P_3+P_6\leq1) \nonumber \\
(q'_0-q'_2) \times (P_0+P_4+P_5+P_6\leq1) \nonumber \\
(q'_7) \times (P_3+P_5+P_6+P_7\leq1) \nonumber \\
\end{eqnarray}

Adding all of these together and we get
\begin{eqnarray}
q'_0P_0+q'_1P_1+q'_2P_2+q'_4P_4+(q'_0+q'_7-q'_4)P_3 \nonumber \\
+(q'_0+q'_7-q'_2)P_5+(q'_0+q'_7-q'_1)P_6+q'_7P_7 \nonumber \\
\leq q'_0+q'_7,
\end{eqnarray}
the same as the previous one.

By using $q'_i+q'_{7-i}=\omega'_c$ we get
\begin{equation}
\omega'_{ns}\leq\omega'_c.
\end{equation}

Putting everything together, we have

\begin{equation}
\max \frac{\omega_{ns}}{\omega_c}=\max (1,\frac{q_0+q_3+q_5+q_6}{3\omega_c},\frac{q_1+q_2+q_4+q_7}{3\omega_c}),
\end{equation}
which completes the proof.

By studying the no-signaling equalities, we will find multipartite
no-signaling correlations that violate the quantum bound.
On the one hand, we may wonder whether no-signaling
theories other than quantum mechanics are necessary and this will motivate us
to explore how much quantum mechanics can be violated by no-signaling correlations, as we have already done.
On the other hand, it is also an interesting question what additional principles we need
to constrain no-signaling theories down to quantum mechanics.

Here, we show explicitly that if we use the Local Orthogonality (LO) restrictions \cite{Fritz},
we will recover the common boundary of classical and quantum mechanics:

\begin{equation}
\max \frac{\omega_{LO}}{\omega_c}=1,
\end{equation}
which means that LO inequalities are complete for this input-undetermined GYNI problem. The proof is straightforward: LO adds two new inequalities, $P_0+P_3+P_5+P_6\leq1$ and $P_1+P_2+P_4+P_7\leq1$ to the list of inequalities. By the previous proof we only need to consider the case $q_0+q_3+q_5+q_6\geq3\omega_c$. Now we set $\min (q_0,q_3,q_5,q_6)=q_0$ without the loss of generality. Then we have

\begin{eqnarray}
\omega_{LO}=\sum_{i=0}^{7}q_iP_i\leq q_0+(q_1P_1+q_2P_2+(q_3-q_0)P_3 \nonumber \\
+q_4P_4+(q_5-q_0)P_5+(q_6-q_0)P_6+q_7P_7).
\end{eqnarray}

Take $(0,q_1,q_2,q_3-q_0,q_4,q_5-q_0,q_6-q_0,q_7)$ as a new input distribution, with classical winning probability $\omega_c-q_0$ and the relationship $q_3+q_5+q_6-3q_0\leq3\omega_c-3q_0=3(\omega_c-q_0)$. Using the same reasoning of the second part of the previous proof, we have

\begin{eqnarray}
q_1P_1+q_2P_2+(q_3-q_0)P_3+q_4P_4 \nonumber \\
+(q_5-q_0)P_5+(q_6-q_0)P_6+q_7P_7\leq\omega_c-q_0,
\end{eqnarray}
which means that $\omega_{LO}\leq\omega_c$.

As we already know that $\omega_c$ is reachable, we conclude that LO will close completely
the gap between no-signaling and quantum. This fact is proven for tripartite state with arbitrary
probability distributions, extending the results of fixed input distributions found in Ref \cite{Fritz}.

\end{document}